\documentclass[11pt,twoside]{article}

\usepackage{asp2006}
\usepackage{epsf}
\usepackage{lscape}

\begin{document}
\title{High-energy flares from jet-clump interactions}   
\author{Anabella T. Araudo$^{1,2}$
Valent{\'\i} Bosch-Ramon$^{3}$ and Gustavo E. Romero$^{1,2}$}   
\affil{$^{1}$ Instituto Argentino de Radioastronom{\'\i}a (CCT La Plata, 
CONICET), C.C.5, 1894 Villa Elisa,  Buenos Aires, Argentina
\\
$^{2}$ Facultad de Ciencias Astron\'omicas y Geof\'{\i}sicas,
Universidad Nacional de La Plata, Argentina 
\\
$^{3}$  Max Planck Institut f\"ur Kernphysik, Saupfercheckweg
1, Heidelberg 69117, Germany} 

\begin{abstract}
High-mass microquasars are binary systems composed by a 
massive star and a compact object from which relativistic jets are
launched.   
Regarding the companion star, observational 
evidence supports the idea 
that winds of hot stars are formed by clumps. Then,
these inhomogeneities may interact with the jets producing a flaring activity. 
In the present contribution we study the interaction between a jet 
and a clump of the stellar wind in a high-mass microquasar. 
This interaction produces a shock in the jet, where particles may be 
accelerated up to relativistic energies.  
We calculate the spectral energy distributions  
of the dominant non-thermal processes: synchrotron 
radiation, inverse Compton scattering, and proton-proton 
collisions. Significant levels of X- and $\gamma$-ray emission
are predicted, with luminosities in
the different domains up to $\sim 10^{34} - 10^{35}$ erg s$^{-1}$
on a timescale of about  $\sim 1$~h.
Finally, jet-clump interactions in high-mass microquasars could be 
detectable at high energies.
These phenomena may be behind the fast 
TeV variability found in some high-mass X-ray binary systems, such as 
Cygnus X-1, LS 5039 and LS I+61 303.
In addition, our model can help to derive
information on the properties of jets and clumpy winds.
\end{abstract}

\section{Introduction}  

The mass loss in massive stars is
thought to take place via supersonic inhomogeneous winds. 
Considerable observational evidence supports the idea that the wind
structure is clumpy (e.g. Owocki \& Cohen 2006), 
although the properties of clumps are not 
well-known as a cosequense of the high spatial resolution necessary for 
the clumps detection. 
Some massive stars are accompanied by a compact object
and present transfer of matter to it forming
an accretion disk and,
in high-mass microquasars (HMMQs), bipolar 
relativistic outflows.

Non-thermal emission has been observed in microquasar jets from radio 
to X-rays, and it is thought that
$\gamma$-rays could be also produced in jets.
Recently, a TeV flare has been detected from the 
HMMQ Cygnus~X-1 (Albert et al. 2007). 
Transient $\gamma$-ray events could have
been detected as well from the high-mass X-ray binaries   
LS~5039 (Aharonian et al. 2005), LS~I+61~303 (Albert et al. 2006)
and perhaps from the HMMQ Cygnus X-3 (ATels 1492, 1547 and 1585).
Some authors have been suggested that this strongly variable $\gamma$-ray 
emission can be produced by the interaction between the jet and the
stellar wind of the companion star (e.g. Owocki et al. 2009, 
Araudo et al. 2009).

In the present contribution we propose a model to explain these 
$\gamma$-ray flares based on the interaction between 
the jets of a HMMQ with wind inhomogeneities. 
The clumps can eventually penetrate in the jet, 
leading to transient non-thermal activity that may
release a significant fraction of the jet kinetic luminosity in the
form of synchrotron, inverse Compton (IC), and proton-proton ($pp$) 
emission.

\section{The scenario}

To study the interaction between a clump of the stellar wind 
and a jet in a HMMQ, we adopt a scenario
with similar characteristics to Cygnus~X-1.
We fix the separation between the
compact object and the massive star to 
$a=3\times10^{12}$~cm.
The stellar mass loss rate is adopted to be $\dot{M}_{\star}=3\times
10^{-6}\,M_{\odot}$~yr$^{-1}$, with a terminal wind velocity  
$v_{\rm w} \sim 2.5\times10^8$~cm~s$^{-1}$.
A sketch of the scenario is presented in Figure \ref{Sketch}.

\subsection{Clump and jet model}

The clump is taken to be spherical and we consider two values of it
radius: $R_{\rm c}= 10^{10}$ and $10^{11}$~cm.
We have assumed that the velocity of the clumps is equal to the velocity 
of the wind, i.e $v_{\rm c} = v_{\rm w}$. 
In order to obtain a large density contrast between the clump
and the jet we assume dense and homogeneous clumps with 
$n_{\rm c} = 10^{12}$~cm$^{-3}$, which corresponds to a clumpy wind 
filling factor 
of $f=\dot{M}_{\star}/4\pi a^2 m_p v_{\rm c} n_{\rm c} = 0.005$, 
where $m_p$ is the mass of the proton.

We consider a jet dynamically dominated by cold protons
with a bulk velocity $v_{\rm j} = 0.3\,c$ and
a kinetic luminosity $L_{\rm kin} \sim 3\times10^{36}\;\rm{erg\;s^{-1}}$ 
(e.g. Gallo et al. 2005).
We assume here that the jet radius is one tenth of the jet 
height, i.e. $R_{\rm j}(z) = 0.1\,z$, 
and the jet-clump interaction is taken to occur at $z_{\rm int}=a/2$. 
The density of the jet material at $z_{\rm int}$ results
$n_{\rm j} = 4.7\times10^7$~cm$^{-3}$ and then,
the ratio between the clump and the jet densities is $\chi=2.1\times10^4$. 
This parameter
will be very relevant for the jet-clump interaction estimates.

\subsection{Dynamics of the interaction}

In order to study the physical processes of the jet-clump interaction, 
we consider the collision of  a single clump with the jet.
The huge density contrast $\chi$ allows the clump to cross the
boundary of the jet and fully penetrate into it. 
In the context of this work, thermal conduction, clump expansion, magnetic
fields and gravitational forces are not dynamically relevant for the
jet-clump interaction and will be
neglected. 

In the Laboratory Reference Frame (LRF), the clump will take a time 
$t_{\rm c}$ to fully enter the jet and a time $t_{\rm j}$ to cross it
roughly at the wind velocity: 

\begin{equation} 
t_{\rm c}\sim 2\,R_{\rm c}/v_{\rm c} \sim 80~\rm s\;(R_{\rm c} = 10^{10}~\rm{cm})\; 
\rm{and}
\;\sim 800~\rm s\;(R_{\rm c} = 10^{11}~\rm{cm}),
\end{equation}
\begin{equation} 
t_{\rm j}\sim \frac{2\; R_{\rm j}}{v_{\rm c}} = 1.2\times10^3~\rm s.
\end{equation} 

When the clump interacts with the jet, a shock is formed
in the clump and propagates in the direction of the jet
motion with a velocity $v_{\rm cs}\sim v_{\rm j}/\sqrt{\chi}$. 
The clump-crossing time, which is
the characteristic timescale of the jet-clump interaction,
can be defined as:
\begin{equation} 
t_{\rm cc}\sim \frac{2\; R_{\rm c}}{v_{\rm cs}} \sim 
3\times10^2~\rm s\;(R_{\rm c} = 10^{10}~\rm{cm})\; \rm{and}
\;\sim 3\times10^3~\rm s\;(R_{\rm c} = 10^{11}~\rm{cm}).
\end{equation} 

A shock (the bow shock) is also formed in the jet when its material 
collides with the clump. 
We assume the clump/bow-shock separation
distance as $x\sim 0.2 R_{\rm c}$ (van Dyke \& Gordon 1959), 
and thus the time 
required  to reach the steady state regime results:
\begin{equation} 
t_{\rm bs}\sim\frac{0.2 R_{\rm c}}{v_{\rm j~ps}} \sim 
\frac{0.2 R_{\rm c}}{v_{\rm j}/4} \sim \frac{5}{2}\frac{t_{\rm cc}}{\sqrt{\chi}}
\ll t_{\rm cc}.
\end{equation}  
Once the clump is inside the jet, the latter 
accelerates the clump up to the background velocity, $v_{\rm j}$, with
an acceleration $g\sim v_{\rm j}^2/\chi R_{\rm c}$ 
(e.g. Fragile et al. 2004) in a time
\begin{equation} 
t_{\rm g}\sim\frac{v_{\rm j}}{g} \sim \sqrt\chi\; t_{\rm cc} \gg t_{\rm cc}.
\end{equation}
When one fluid exerts a force against another fluid of different density,
the hydrodynamical Rayleigh-Taylor (RT) instability eventually develops,
leading to the perturbation and potential disruption of the clump.  
On the other hand, after the bow shock is formed, the jet material 
surrounds the clump and
we have two fluids with a large relative velocity (in our case 
$\sim v_{\rm j}$). This situation leads to
Kelvin-Helmholtz (KH) instabilities.
The timescale for these instabilities, considering that the length of the 
perturbed region is $\sim R_{\rm c}$, are:
\begin{equation} 
t_{\rm RT}\sim \sqrt{R_{\rm c}/g} \sim t_{\rm cc}\qquad\rm{and}\qquad
t_{\rm KH}\sim \frac{R_{\rm c}\;\sqrt{\chi}}{v_{\rm j}} \sim t_{\rm cc}.  
\end{equation}   

\begin{figure}[!h]
\plottwo{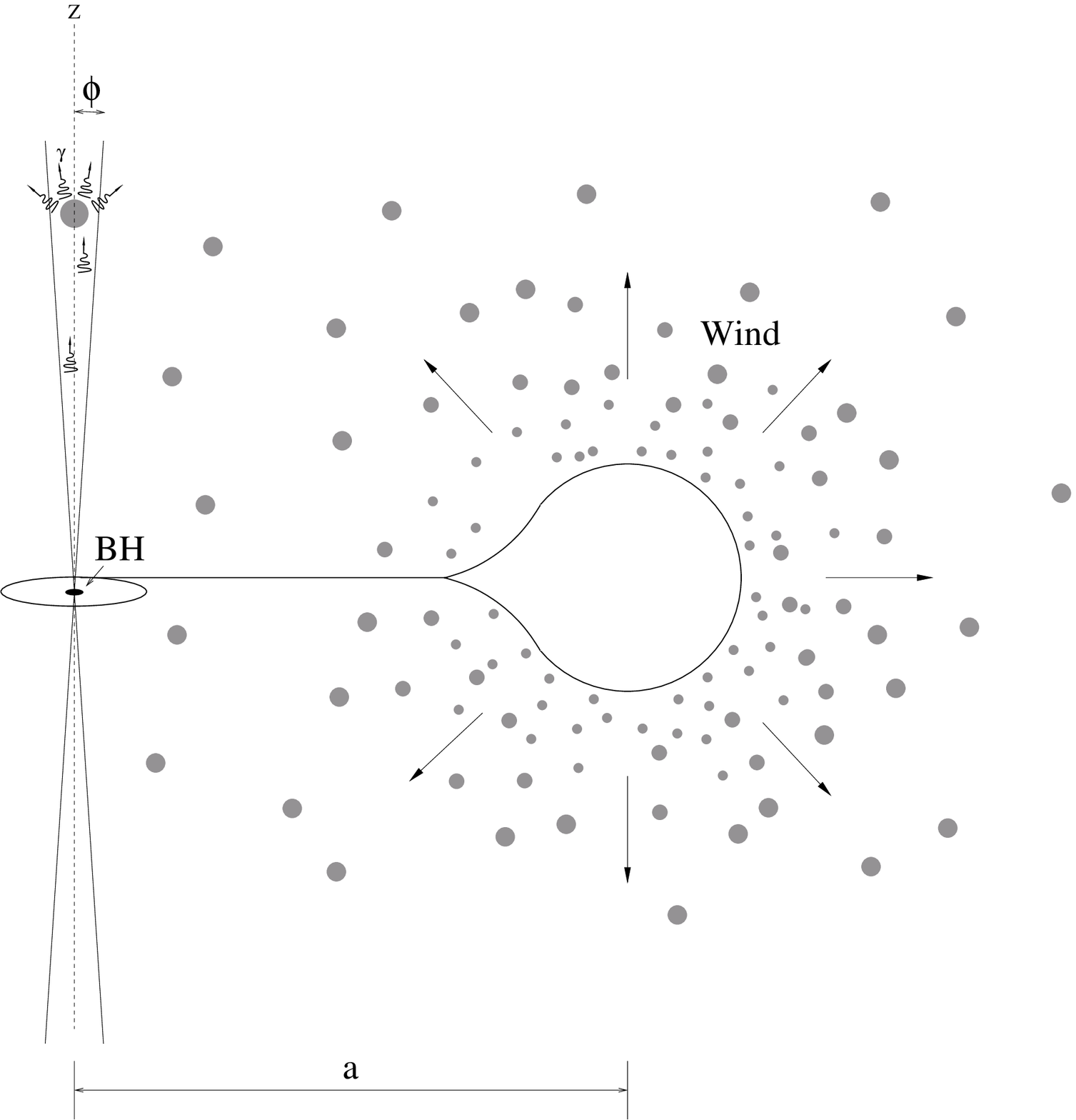}{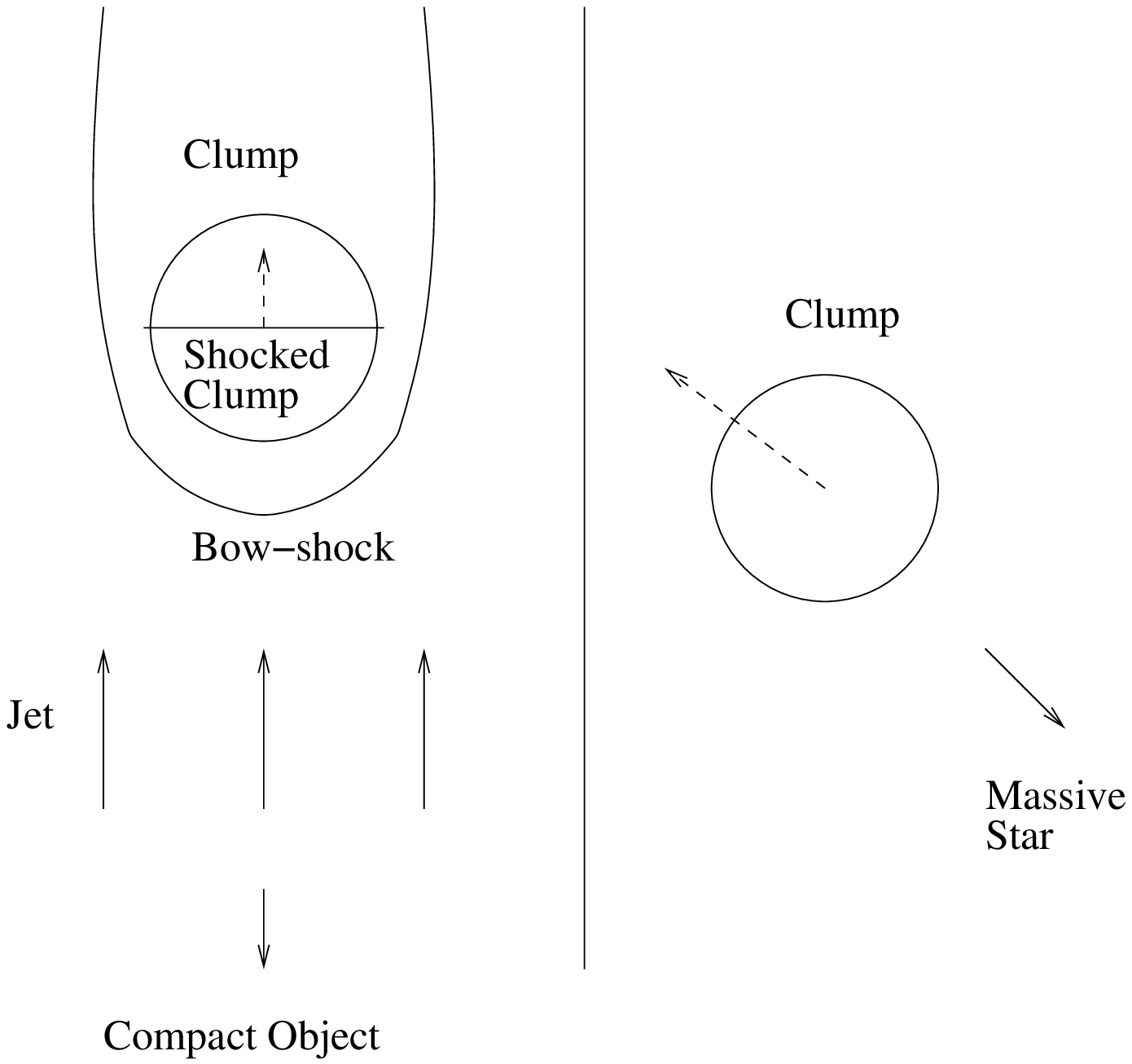}
\caption{Sketch of a HMMQ with a clumpy wind (left) and the jet-clump 
interaction (right).
}\label{Sketch}
\end{figure}

We assume that for a $t_{\rm c} \sim t_{\rm cc}/5$ we can treat the jet-clump 
interaction as the clump were fully inside the jet.
At this stage we do not consider the penetration of the clump into the jet,
and assume in our treatment that the former is completely inside the 
latter (i.e. the system 
has cylindrical symmetry; see Fig. \ref{Sketch}).   
We notice that the clump  might not escape the jet if
$t_{\rm j} > t_{\rm RT/KH}$ due to disruption through
instabilities, although
numerical simulations show that  
the instability timescales are actually longer than the 
clump crossing time by a factor of a few ( e.g. Klein, McKee \& 
Colella 1994). 

\section{Non-thermal processes}

Given the characteristics of our scenario, the shock in the clump 
is radiative, and the bow shock is adiabatic. 
For these reasons, the shocked and heated material of the clump will 
radiate (by thermal Bremsstrahlung)
a non-negligible part of the energy tranferred by the jet, whereas in the
bow shock non-thermal processes take place. We focus here in the non-thermal
emission.

\subsection{Particle acceleration and radiative cooling}
\label{part_accel}

Assuming that non-relativistic diffusive (Fermi I) shock acceleration 
takes place in the bow-shock,  
electrons and protons with a charge $q$ will be accelerated 
up to an energy $E_{e,p}$ in a time 
\begin{equation} 
t_{\rm acc} =  \frac{8}{3}\left(\frac{c}{v_{\rm j}}\right)^2 
\frac{E_{e,p}}{q\,B_{\rm bs}\,c}\,,
\label{t_acc}
\end{equation} 
in the Bohm limit and for perpendicular shocks (Protheroe 1999).
The magnetic field in the acceleration region is $B_{\rm bs}$,  
and we will consider two values for it.
First, we consider a value obtained assuming
that the magnetic energy density is a $10\,\%$
of the plasma internal energy density downstream and we obtain  
$B_{\rm bs} = 150$~G. We adopt a sub-equipartition 
value to make $B_{\rm bs}$ dynamically negligible with respect to matter. 
In addition, we adopt $B_{\rm bs} = 1$~G  to check the impact on
our results of a magnetic field much smaller than in the 
sub-equipartition case.

In the scenario presented in this work, relativistic leptons lose their
energy mainly  by synchrotron radiation and IC scattering. 
The former have the following cooling time
\begin{equation} 
t_{\rm syn} =  \frac{4.1\times10^2}{B_{\rm bs}^2\,E_e}~\rm{s}\,.
\end{equation} 
At $z_{\rm int}$, the energy density of 
the photons from the star is 
$u_{\rm ph} = 2.4\times10^{2}$~erg cm$^{-3}$, being the typical photon energy
$\epsilon_0 \sim 10$~eV. For $y = \epsilon_0 E_e/(5.1\times10^5\, 
\rm{eV})^2 > 1$ 
the IC interaction takes place in the Klein-Nishina (KN) regime. 
A formula for the IC cooling time valid in both Thompson (Th) and
KN regime under a photon field with a narrow energy distribution 
is (e.g. Bosch-Ramon \& Khangulyan 2009):
\begin{equation} 
t_{\rm IC} =  \frac{6.1\times10^{12}\,\epsilon_0}{u_{\rm ph}}
\frac{(1 + 8.3y)}{\ln(1+0.2\;y)}\frac{(1 + 1.3y^2)}{(1 + 0.5y + 1.3y^2)}
~\rm{s}\,.
\end{equation} 
 
Regarding hadronic emission, $\gamma$-rays are produced if 
relativistic protons interact with nuclei through inelastic $pp$ 
collisions, beeing the proton cooling time 
\begin{equation} 
t_{pp} = \frac{2.2\times10^{15}}{4 n_{\rm j}\,\left(0.95 + 0.06\ln\left(\frac{E_p}
{1.1 m_pc^2}\right)\right)}~\rm{s}.
\end{equation} 
In the bow-shock region, this process is negligible. Otherwise, if
relativistic protons accelerated in the bow shock penetrate into the clump,
$pp$ interactions can become an efficient process to generate
$\gamma$-rays because $n_{\rm c} \gg n_{\rm j}$.

\begin{figure}[!h]
\plottwo{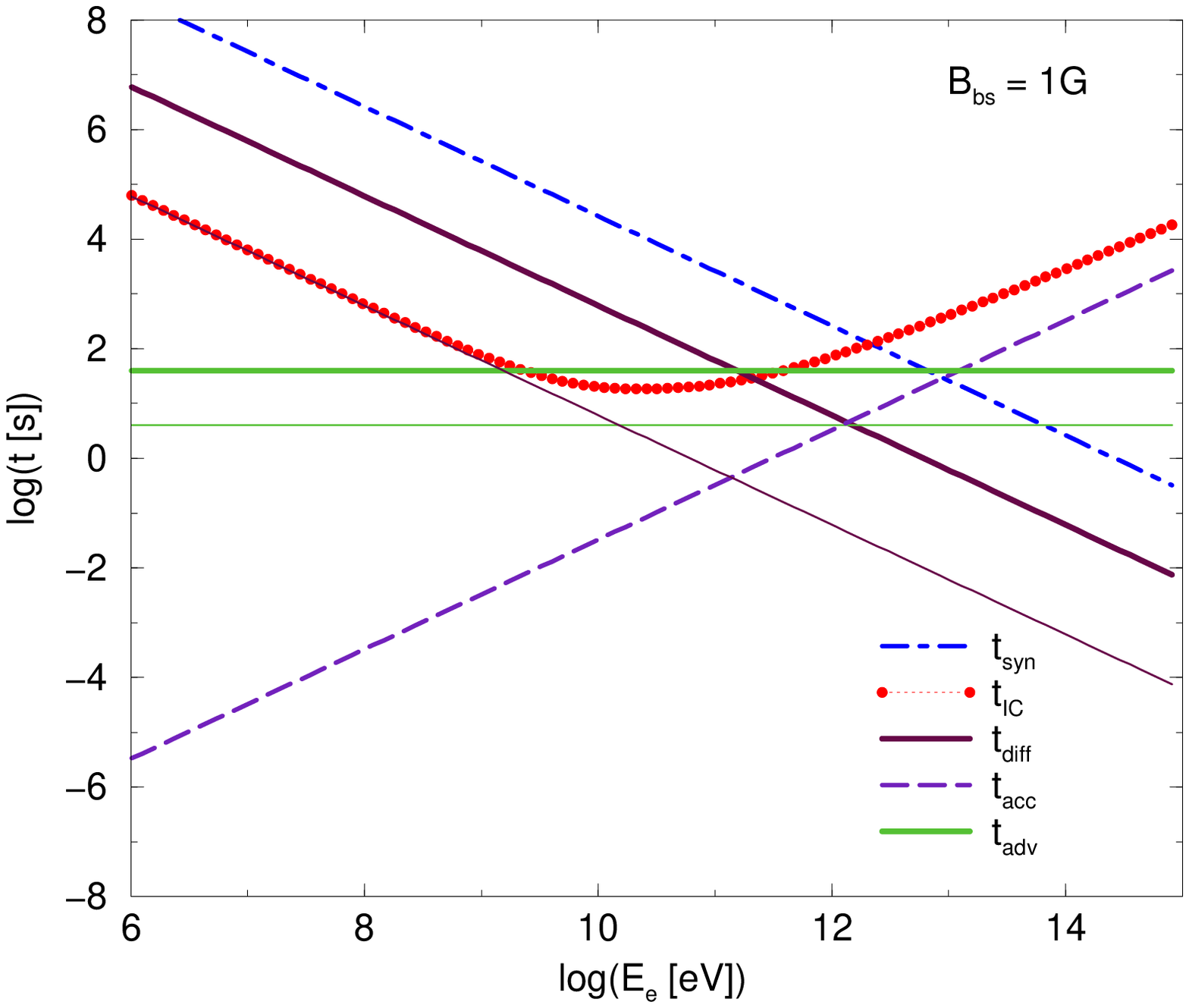}{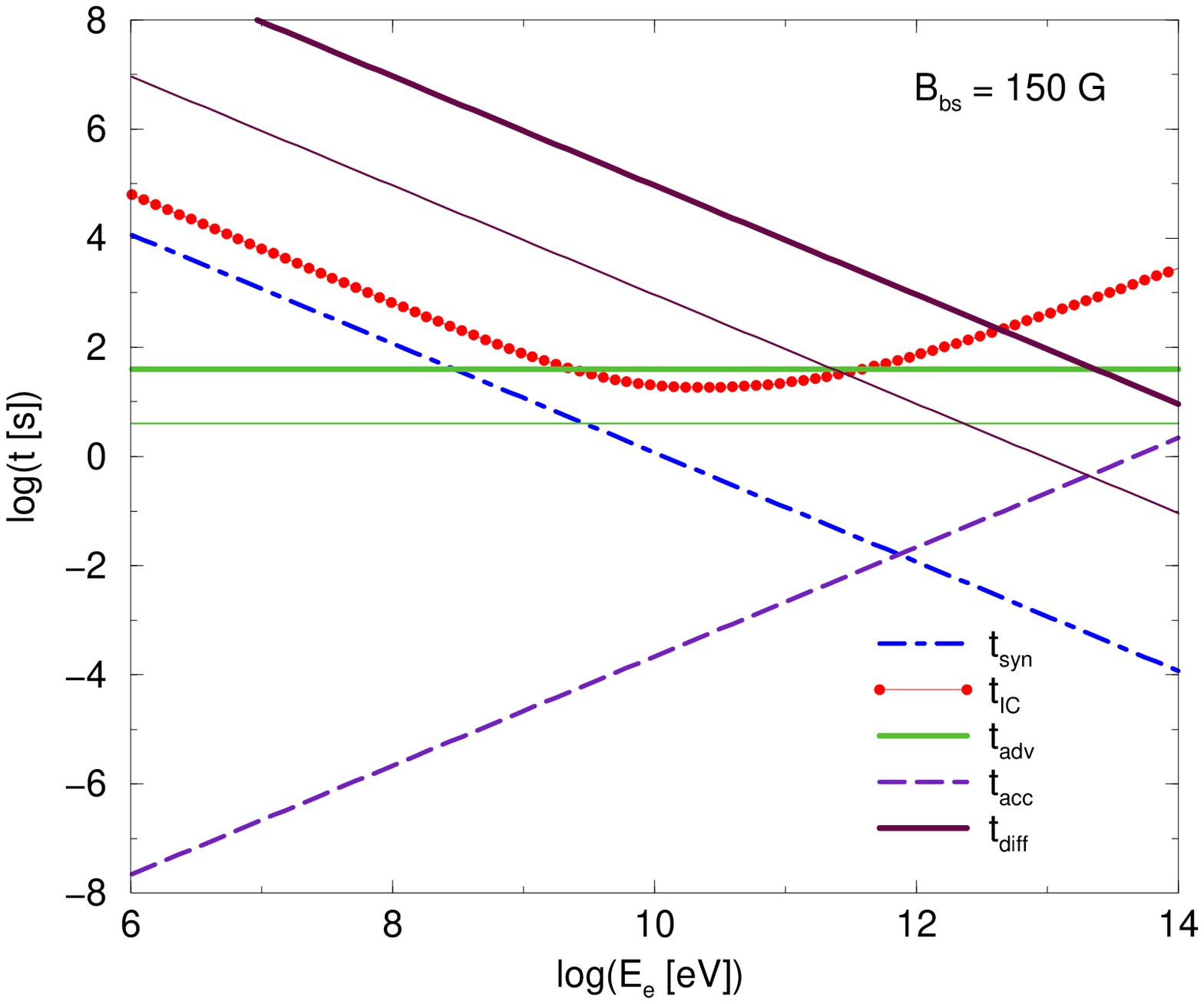}
\caption{
Acceleration and radiative loss time for 
electrons in the bow-shock region. The advection and diffusion times are 
shown for $R_{\rm c} = 10^{10}$ (thin line) and $10^{11}$~cm
(thick line). 
}\label{losses}
\end{figure}

Besides radiative cooling, the escape of particles from
the acceleration region is another kind of energy losses.
The corresponding timescale, $\tau_{\rm esc}$, 
takes into account the advection of relativistic particles by downstream
bow-shock material ($t_{\rm adv} \sim R_{\rm c}/v_{\rm j\,ps}$)
and the diffusion of particles 
($t_{\rm diff} \sim x^2/2 D_{\rm B}$, where 
$D_{\rm B}$ is the Bohm diffusion coefficient), being
$\tau_{\rm esc} = \rm{min}\{t_{\rm diff}, t_{\rm adv}\}$.
Equating energy gains and losses, the maximum energy achieved by
electrons ($E_e^{\rm max}$) accelerated in the bow shock is estimated 
(see Fig. \ref{losses}). 
On the other hand, the maximum energy for protons ($E_p^{\rm max}$)  
is constrained by the size of the acceleration region. 
In Table \ref{Table_energies}, $E_e^{\rm max}$ and $E_p^{\rm max}$
are shown.

\begin{table}[!ht]
\caption{Maximum energies achieved by acelerated particles in the bow shock
and the energy break of electrons.}\label{Table_energies}
\begin{center}
\begin{tabular}{c|cccc}
\hline 
$R_{\rm c}$ [cm] & $10^{10}$ & $10^{10}$ & $10^{11}$ & $10^{11}$ \\
$B_{\rm bs}$ [G] & $1$ & $150$ & $1$ & $150$ \\ 
\hline 
$E_e^{\rm max}$ [eV] & $1.5\times 10^{11}$ & $8\times 10^{11}$ & 
$1.5\times10^{12}$ & $8\times 10^{11}$  \\  
$E_p^{\rm max}$ [eV]& $6\times 10^{11}$ & $9\times 10^{13}$ & 
$6\times10^{12}$ & $9\times 10^{14}$  \\
$E_{\rm b}$ [eV] & $4.5\times 10^{10}$ & $3\times 10^{9}$ & 
$4.5\times 10^{11}$ & $3\times 10^{8}$ \\
\hline
\end{tabular}
\end{center}
\end{table}

\subsection{High energy emission}
\label{gamma}

In this section we calculate the Spectral Energy Distribution (SED)
of the emission produced by the most relevant non-thermal radiative 
processes: synchrotron and IC radiation in the bow-shock region and
$pp$ in the clump.

We assume an injected population of relativistic particles 
in the bow-shock region
that follows a power-law energy distribution of the form:
\begin{equation} 
Q_{e,p}(E_{e,p}) = K_{e,p}\,E_{e,p}^{-2}\,
\exp(-E_{e,p}/E_{e,p}^{\rm max})\;,
\end{equation}
typical for Fermi I acceleration.
The normalization constant $K_{e,p}$ is determined assuming that 
the (non-thermal) luminosity of accelerated particles
is $L_{\rm NT} = 0.25 (R_{\rm c}/R_{\rm j})^2L_{\rm kin}$
(note that the predicted fluxes 
scale linearly with the adopted non-thermal fraction $0.25$).
To estimate the particle energy distribution $N_{e,p}$, we solve
the kinetic equation in the one-zone model approximation for the bow-shock
region (e.g. Ginzburg \& Syrovatskii 1964):

\begin{equation} 
\frac{\partial N_{e,p}}{\partial t} + \frac{\partial}{\partial E_{e,p}}
(\frac{dE_{e,p}}{dt} N_{e,p}) +
\frac{N_{e,p}}{\tau_{\rm esc}} = Q_{e,p}\,.
\end{equation}

The relativistic leptons reach the steady state well before the 
shock has crossed the clump. 
Particles downstream with low energies escape
advected in the shocked material of the jet before cooling radiatively, 
producing a break in the energy spectrum of particles. 
From $t_{\rm adv} = t_{\rm syn}$ and $t_{\rm adv} = t_{\rm diff}$ we determine
the electron break energies $E_{\rm b}$.
The values are shown in Table \ref{Table_energies}.
The most energetic electrons can diffuse up to the
clump ($B_{\rm bs} = 1$~G) or lose their energy
inside the bow-shock region by synchrotron and IC radiation 
($B_{\rm bs} = 150$~G). 
We note that
the electrons with energies $E_e > E_{\rm b}$ can reach
the clump and radiate inside it. In this work, we do not estimate
the lepton emission in the clump.

As noted in Sect. \ref{part_accel}, relativistic protons do not suffer
significant $pp$ losses in the bow-shock region ($t_{\rm diff} \ll t_{pp}$).
These protons can also reach the clump if they are not
advected by the shocked 
material of the jet. Imposing that $t_{\rm diff} < t_{\rm adv}$,
the minimum energy necessary to reach the clump is
$E_p^{\rm min} = 0.025 E_p^{\rm max}$,
and the  maximum energies of these protons 
are shown in Table \ref{Table_energies}.
On the other hand, to confine relativistic protons in the clump  
requires a very large magnetic field
and we assume that protons 
will cross the whole clump in a time 
$\sim R_{\rm c}/c < t_{pp}$. 
Then, the distribution of relativistic protons in the clump is
$N_p(E_p) = (R_{\rm c}/c) Q_p(E_p)$.

With the steady distributions of relativistic electrons in the
bow shock, $N_e(E_e)$, and protons in the clump, $N_p(E_p)$, 
we calculate the SEDs.
We estimate the specific luminosity as
\begin{equation}
\epsilon L(\epsilon) = 
\epsilon \int_{E_{e,p}^{\rm min}}^{E_{e,p}^{\rm max}} N_{e,p}(E_{e,p})\;
j(E_{e,p}, \epsilon)\,dE_{e,p}\,,
\end{equation}
where $j(E_e, \epsilon)$ is the emissivity of the process
 and $\epsilon$ is the photon energy.

Synchrotron and IC emissivities are computed using the standard 
formulae given in Blumenthal \& Gould (1970). For the later process
we consider 
the parametrized differential cross section valid in both Th and KN regimes
and $\pi^0$-decay emissivity is calculated following Kelner et al. (2006).
The synchrotron emission is computed in the optically thin case,
neglecting the impact of synchrotron self-absorption, but
at high energies (IC and $pp$) $\gamma-\gamma$ absorption is taken into
account.

\begin{figure}[!ht]
\begin{center}
\plotone{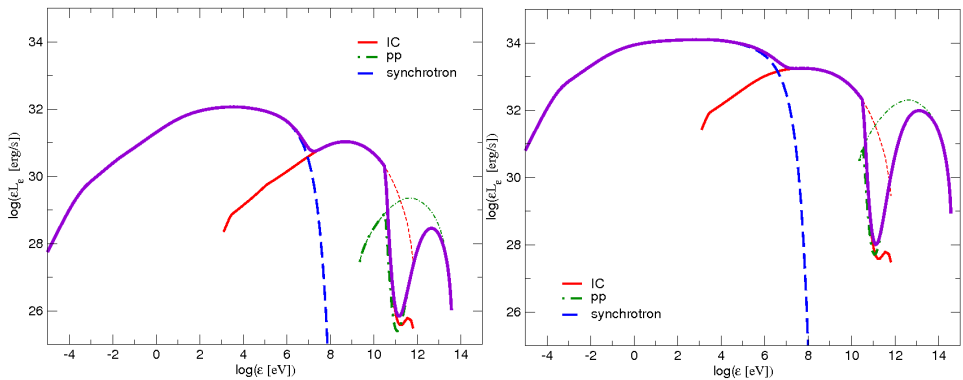}
\caption{Leptonic and hadronic
emission from the bow shock and from the clump, respectively, for 
$B_{\rm bs} = 150$~G, and $R_{\rm c}=10^{10}$ (left) and $10^{11}$~cm (right).
The curves of both absorbed (thick lines) and unabsorbed (thin lines) 
IC and $pp$ radiation are shown.}
\label{SEDs}
\end{center}
\end{figure}

Synchrotron emission  is dominant at X-rays for $B_{\rm bs}=150$~G,
with $L_{\rm syn} \sim 10^{35}$~erg~s$^{-1}$.
Inverse Compton scattering produces $\gamma$-rays up to 
very high energies, dominating the radiation output in the cases with
$B_{\rm bs}=1$~G. In our
calculations, the highest luminosity achieved is $L_{\rm IC}\sim
10^{35}$~erg~s$^{-1}$ for $R_{\rm c}=10^{11}$~cm, although
$\gamma-\gamma$ absorption can reduce substantially the emission above
100~GeV. Proton-proton collisions in the clump can also produce 
$\gamma$-rays at energies that may be as high as $\sim
10^{14}$~eV ($B_{\rm bs}=150$~G).
The maximum luminosity obtained by
$pp$ is nevertheless quite modest, $L_{pp} \sim 10^{32}$~erg~s$^{-1}$
for $R_{\rm c}=10^{11}$~cm, although denser and/or bigger clumps, and
more powerful jets may yield detectable amounts of photons outside 
the $\gamma-\gamma$ absorption range (0.1-10~TeV). 
The computed SEDs are shown in Fig. \ref{SEDs}, only for the cases
with $B_{\rm bs} = 150$~G. 


\section{Effect of many clumps}

In general, many clumps can interact with the jet and is possible to
have several clumps inside it simultaneously (see Owocki et al. 2009). 
However, at $z < z_{\rm c} = 4\times10^{11}$~cm the destruction 
time $t_{\rm cc}$ is less than the penetration one ($t_{\rm c}$) and
the clump will be destroyed before completely entering into the jet. 
Then, the model presented here is not valid at $z < z_{\rm c}$ 
(neither if $R_{\rm j} < R_{\rm c}$).

Considering a canonical geometry for the jet and a filling factor of the wind
$f = 0.005$, we estimate the number of 
clumps -$N_{\rm c}$- inside the jet, from $z = z_{\rm c}$ up to $z \sim a$. 
We obtain
$N_{\rm c} \sim 350$ and $\sim 0.5$ for $R_{\rm c}=10^{10}$ and 
$10^{11}$~cm, respectively. As a consecuence,
flares produced by jet-clump interactions 
could be a sporadic phenomenon, for a low $N_{\rm c}$
($R_{\rm c}=10^{11}$~cm), 
or may appear as a modulated steady activity, for a high $N_{\rm c}$
($R_{\rm c}=10^{10}$~cm). 
In the latter case, the resulting SED will be similar than
the obtained in the previous section, but multiplied by $N_{\rm c}$.
Nevertheless, we note that
the jet may be disrupted in those cases when too many clumps are 
simultaneously present inside the jet (Araudo et al. 2009). 
However, more detailed calculations of the dynamics of the jet-clump 
interaction are required to clarify this issue.

\section{Conclusions}

In the present work, we study the main physical processes and the
radiative outcomes from the interaction between the jet of a HMMQ with a
clump of the companion stellar wind. 
This interaction produces a bow-shock in the jet where particles are
accelerated up to very-high energies. The relativistic electrons are 
cooled efficiently by synchrotron and IC radiation in the bow shock, 
and protons by $\pi^0$-decay from $pp$ interactions in the clump.  

Due to characteristics of the interaction, the expected
emission can be transient if $N_{\rm c}$ is low. Given the typical dynamical 
timescales, the flare duration may last from minutes to a few hours.
The flares of jet-clump interactions would have associated lower
(synchrotron) and higher energy components (IC, $pp$ emission).
The total level of emission,
the importance of the different components, and the duration of the
flares, can give information on the jet power and size, clump size and
density, and magnetic fields in the interaction regions. Therefore,
besides the jet itself, the clump properties can be probed by
observations at high energies of transient activity in HMMQs, 
being a new tool to study the winds of massive stars.

\acknowledgements 
The authors thank S. Owocki and D. Khangulyan for many insightful 
discussions.
A.T.A. thanks the Max Planck Institut f$\rm\ddot{u}$r Kernphysik for 
its suport and kind hospitality.
V.B-R. and G.E.R. acknowledge support by DGI
of MEC under grant AYA2007-68034-C03-01, as well as partial support by
the European Regional Development Fund (ERDF/FEDER).
V.B-R. gratefully acknowledges support from the Alexander von Humboldt
Foundation.

\end{document}